\documentclass[aps,prl,reprint,superscriptaddress]{revtex4-1}
\usepackage{graphicx}
\usepackage{lipsum}
\usepackage{nicefrac}
\usepackage{txfonts}
\usepackage{bm}
\usepackage{epstopdf}
\usepackage{placeins}
\usepackage{units}
\usepackage{color}
\usepackage{soul}
\begin{document}

\title{Slow noncollinear Coulomb scattering in the vicinity of the Dirac point in graphene}

\author{J. C. K\"onig-Otto}
\email[]{j.koenig-otto@hzdr.de}
\affiliation{Helmholtz-Zentrum Dresden-Rossendorf, P.O. Box 510119, 01314 Dresden, Germany}
\affiliation{Technische Universit\"at Dresden, 01062 Dresden, Germany}

\author{M. Mittendorff}
\affiliation{University of Maryland, College Park, MD 20742, USA}

\author{T. Winzer}
\affiliation{Technische Universit\"at Berlin, Hardenbergstra\ss{}e 36, 10623 Berlin, Germany}

\author{F. Kadi}
\affiliation{Technische Universit\"at Berlin, Hardenbergstra\ss{}e 36, 10623 Berlin, Germany}

\author{E. Malic}
\affiliation{Chalmers University of Technology, SE-41296 G\"oteborg, Sweden}

\author{A. Knorr}
\affiliation{Technische Universit\"at Berlin, Hardenbergstra\ss{}e 36, 10623 Berlin, Germany}

\author{C. Berger}
\affiliation{Georgia Institute of Technology, Atlanta, GA 30332, USA}
\affiliation{Institut N\'eel, CNRS-Universit\'e Alpes, 38042, Grenoble, France}

\author{W. A. de Heer}
\affiliation{Georgia Institute of Technology, Atlanta, GA 30332, USA}

\author{A. Pashkin}
\affiliation{Helmholtz-Zentrum Dresden-Rossendorf, P.O. Box 510119, 01314 Dresden, Germany}

\author{H. Schneider}
\affiliation{Helmholtz-Zentrum Dresden-Rossendorf, P.O. Box 510119, 01314 Dresden, Germany}

\author{M. Helm}
\affiliation{Helmholtz-Zentrum Dresden-Rossendorf, P.O. Box 510119, 01314 Dresden, Germany}
\affiliation{Technische Universit\"at Dresden, 01062 Dresden, Germany}

\author{S. Winnerl}
\affiliation{Helmholtz-Zentrum Dresden-Rossendorf, P.O. Box 510119, 01314 Dresden, Germany}

\date{\today}

\begin{abstract}
The Coulomb scattering dynamics in graphene in energetic proximity to the Dirac point is investigated by polarization resolved pump-probe spectroscopy and microscopic theory. Collinear Coulomb scattering rapidly thermalizes the carrier distribution in $\bm{k}$-directions pointing radially away from the Dirac point. Our study reveals, however, that in almost intrinsic graphene full thermalization in all directions relying on noncollinear scattering is much slower. For low photon energies, carrier-optical-phonon processes are strongly suppressed and Coulomb mediated noncollinear scattering is remarkably slow, namely on a ps timescale. This effect is very promising for infrared and THz devices based on hot carrier effects.
\end{abstract}


\maketitle

\begin{figure}[t]
\includegraphics[trim=0.4cm 0.1cm 0.5cm 4.3cm,width=0.48\textwidth]{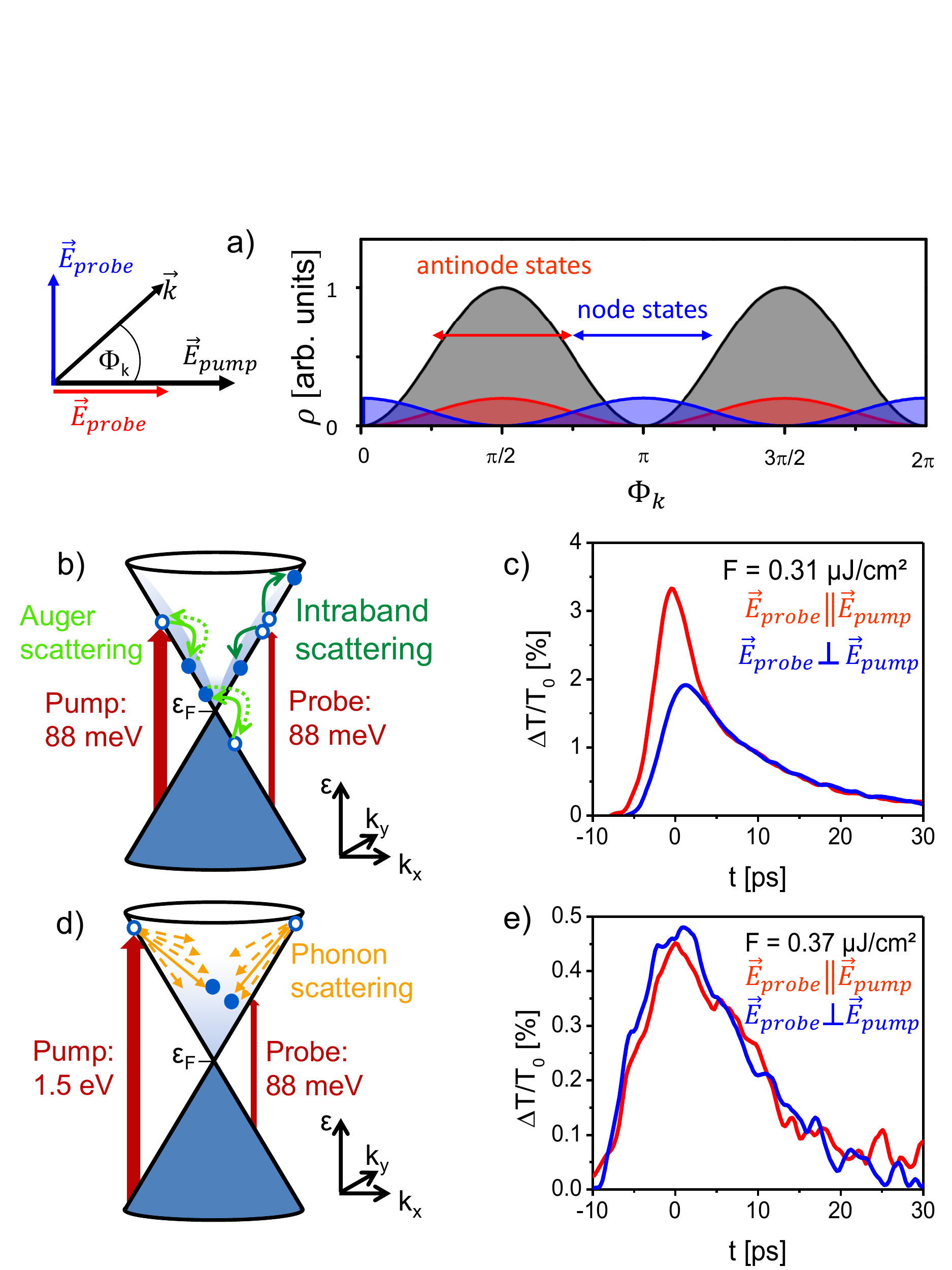}
	\caption{a) Schematic of the angular dependence of the occupation of optically excited carriers in the conduction band at the energy $\hbar \omega/2 = \unit[44]{meV}$ excited by the pump beam (black), a probe beam co-polarized to the pump beam (red), and cross-polarized to the pump beam (blue). Schematic representation of the single-color b) and two-color d) experiments with important scattering mechanisms. Pump-induced transmission change for probing the antinode (red curves) and node states (blue curves), respectively, for the single-color c) and two-color e) experiment.\label{fig:twocol}}
\end{figure}

Coulomb scattering is a nonlinear many-body effect that transforms a non-equilibrium carrier distribution in a semiconductor into a hot Fermi-Dirac distribution \cite{Koch1992}. In the gapless semiconductor graphene with linear energy dispersion Coulomb scattering is known to be particularly strong \cite{MalicPRB2011,Ploetzing2014,Sun2012,Tomadin2013,NPHYMittendorff2014}. A large number of both degenerate and multi-color pump-probe experiments, nonlinear THz spectroscopy, time-resolved photocurrent measurements and time-resolved angle-resolved photoelectron emission spectroscopy (tr-ARPES) have provided detailed insights into the carrier dynamics of graphene \cite{Dawlaty2008,Breusing2011,Obraztsov2011,Tielrooij2013,Mics2015,Jensen2014,TielrooijNNano2015,Johannsen2013,GierzNMat2013}. Together with theoretical studies they revealed that typically carriers thermalize on a sub-100 fs timescale \cite{Breusing2011,Mics2015,Gierz2015}. This rapid transition from a non-equilibrium distribution to a hot thermalized distribution is mediated by both Coulomb scattering and scattering via optical phonons. These two processes have so far not been disentangled as they occur typically on the same time and energy scales \cite{Breusing2011,Jensen2014,Gierz2014}. However, when not just the electron energy but also the angular distribution in $\bm{k}$-space is considered, it has been found that the distribution, in fact, thermalizes rapidly along all $\bm{k}$-directions pointing radially away from the Dirac point. In contrast, the thermalization between different $\bm{k}$-directions is considerably slower. This angular thermalization that is mediated mainly by efficient optical phonon scattering is completed only after 150\,fs \cite{Mittendorff2014,Yan2014,Trushin2015,Malic2012,Satou2015}. In these studies an initial anisotropic electron distribution is generated by pumping with linearly polarized radiation \cite{Gruneis2003,MalicPRB2011,Malic2012}. This anisotropy is linked to pseudospin flipping in interband transitions \cite{Trushin2011,Echtermeyer2014}.

In this Letter, we utilize this anisotropy to trace the temporal and directional characteristics of pure Coulomb scattering in the vicinity of the Dirac point. To this end, we apply a photon energy of 88 meV that is well below the optical phonon energy ($\sim$200\,meV) so that scattering via optical phonons is strongly supressed. In order to ensure that also scattering of the thermalized carrier distribution with optical phonons is negligible, the study is carried out at low temperature and low enough fluences. Acoustic phonon scattering takes place on a longer, \unit[100]{ps} timescale. Both, the experiments and the theory, reveal that in this regime angular thermalization characterized by an isotropic distribution is approached remarkably slowly, namely on a ps timescale. 

\begin{figure}[tbp]
\includegraphics[width=0.48\textwidth]{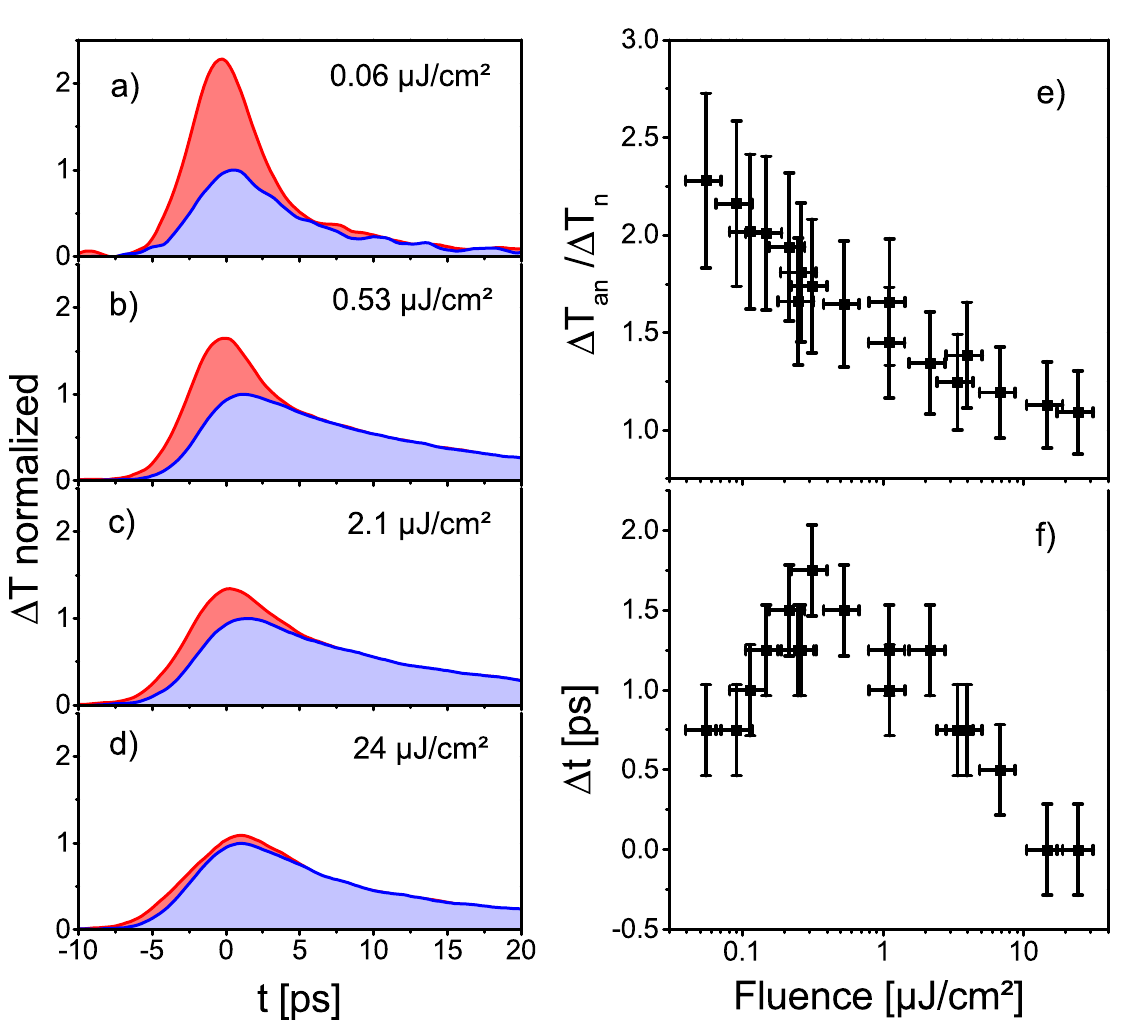}
	\caption{a)-d) Single-color differential transmission for probing the antinode (red curves) and node (blue curves) states, respectively, normalized to the peak value obtained for probing the node states for different pump fluences. e) Fluence dependence of the maximal induced transmission change for probing the antinode states divided by the maximal induced transmission change for probing the node states. f) Fluence dependence of the temporal shift of the maximum induced transmission for probing the node states with respect to the maximum obtained for probing the antinode states. \label{fig:pp}}
\end{figure}

\begin{figure}[t]
\includegraphics[width=0.48\textwidth]{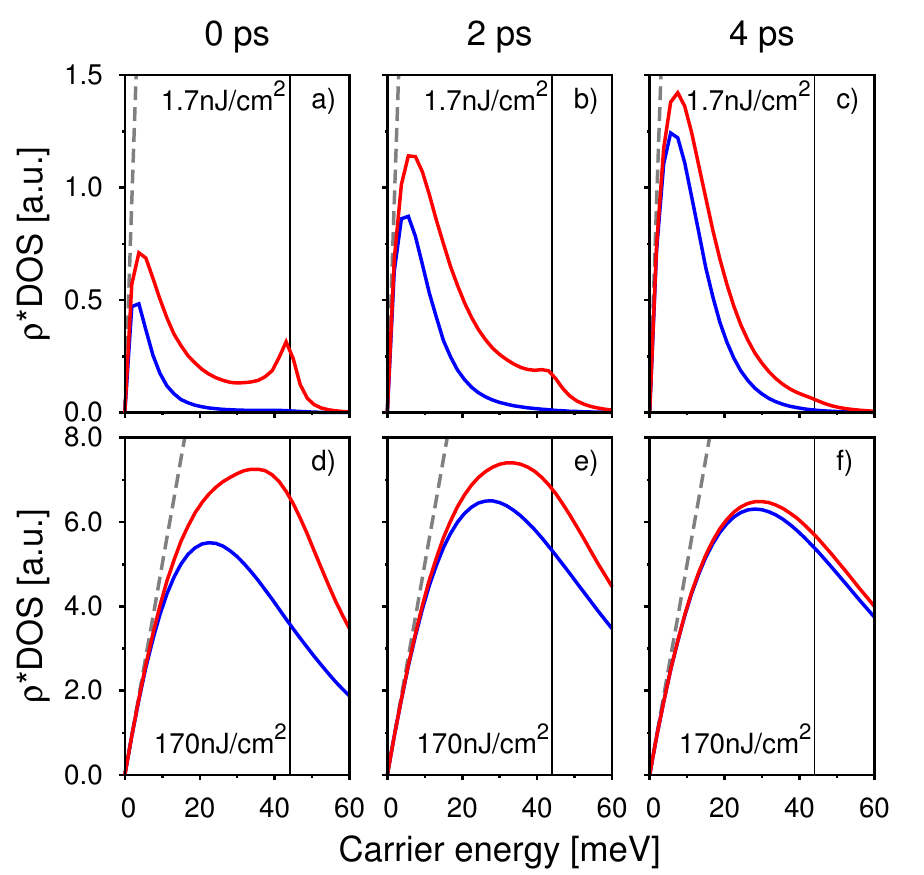}
	\caption{Carrier occupation multiplied by the density of states for three different times (time zero corresponds to maximum intensity of the pump pulse) and two different pump fluences. Red curves correspond to the $\bm{k}$-direction $\bm{k} \bot \bm{E_{pump}}$ ($\Phi_k=\pi/2$), blue curves to $\bm{k} \| \bm{E_{pump}}$ ($\Phi_{k} = 0$). The solid line denotes the energy $\hbar\omega/2$, the dashed line half-filling of the available states. \label{fig:theory}}
\end{figure}
The pump-probe experiments presented here are carried out in transmission geometry on epitaxial multilayer graphene ($\sim$50 layers), which was grown on the C-face of 4H-SiC. Explicitly the layers are rotationally stacked, so that each layer has the electronic structure of graphene \cite{Berger2006,Sprinkle2009}. The sample was kept at 20\,K and differential transmission signals (DTSs) were recorded. In our degenerate pump-probe experiments the free-electron laser FELBE provides radiation at a photon energy of 88\,meV (pulse duration 4\,ps, repetition rate 13\,MHz). At this photon energy the measured signals stem predominantly from interband transitions in the almost intrinsic graphene layers ($|E_{f}|=10-20$\,meV), and only negligible contributions from intraband absorption are expected \cite{Sun2008,Winnerl2011,Kadi2014}. The probe-beam polarization is set to $45^{\circ}$ with respect to the polarization of the pump beam by a grating polarizer on a polymer foil. Polarizers mounted in front of a mercury cadmium telluride (MCT) detector with orientation parallel or perpendicular to the pump beam polarization, allow one to measure the pump-induced transmission for $\bm{E_{probe}} \| \bm{E_{pump}}$ and $\bm{E_{probe}} \bot \bm{E_{pump}}$, respectively. Normal incidence on the sample is used for the probe beam, while the pump beam is offset with respect to the probe beam by a small angle. We also performed a two-color experiment, where a Ti:sapphire laser synchronized to FELBE \cite{Bhattacharyya2011} with a jitter of 5\,ps is employed for pumping at a photon energy of 1.5\,eV (pulse duration 3\,ps, repetition rate 78\,MHz) while probing at 88\,meV. Also in this experiment the response is dominated by interband transistions in the almost intrinsic graphene layers.

In our theoretical investigation we solve the graphene Bloch equations based on the Heisenberg equation of motion in Born-Markov approximation \cite{lindberg88,Knorr96,Malic13}. By accounting for the semi-classical light-matter, carrier-carrier, and carrier-phonon interactions on a consistent microscopic footing we are able to realistically model the carrier dynamics resolved in time, energy, and momentum angle without any adjustable parameter \cite{MalicPRB2011}. The screening of the Coulomb interaction is treated within the static limit of the Lindhard equation \cite{Koch1992}. Dynamical screening \cite{Tomadin2013,Kadi15} yields very similar results due to the induced broadening of the phase space for collinear scattering and the resulting softening of the strict energy conservation \cite{Winzer13}. The latter has been included in a self-consistent procedure taking into account the finite lifetime of two-particle correlations for electrons or phonons \cite{Schilp94,Winzer13}.

In Figure 1a we show schematically the expected results for the angle-dependence of the optically induced population $\rho$ without scattering or saturation effects. It shows a clear anisotropy and follows a $\rho \sim |\sin(\Phi_{k})|^2$ dependence featuring nodes at $\Phi_{k} = 0$ (and $\pi$), where $\Phi_{k}$ is the angle between the amplitude of the pump beam $\bm{E_{pump}}$ and the $\bm k$-vector of electrons (defined as the relative momentum with respect to the Dirac point) \cite{MalicPRB2011}. In the following, the states around $\Phi_{k} = \pi/2$ (and $3\pi/2$) with respect to the pump beam will be referred to as antinode states and the states around $\Phi_{k} = 0$ (and $\pi$) as node states. Due to Pauli blocking the $\bm{E_{probe}} \| \bm{E_{pump}}$ ($\bm{E_{probe}} \bot \bm{E_{pump}}$) configuration preferentially probes the antinode (node) states. Before discussing the experimental findings of the degenerate pump-probe experiments in detail, the main aspects of the involved physics are briefly highlighted. For energies below the optical phonon energy, Coulomb interaction is expected to be the main scattering mechanism. This involves both intraband and interband (Auger-type) processes (cf. Fig. 1b). Coulomb scattering in graphene is predominantly collinear, as $|V|^2\propto(1+\cos\phi_1)(1+\cos\phi_2)$, where $V$ is the intraband Coulomb matrix element and $\phi_i$ is the angle between the momenta of initial and final state of two scattering electrons $i=1,2$ \cite{Malic2012,Song2013}. That means the scattering cross section is maximal for collinear processes ($\phi_i=0$), decreases with increasing scattering angle and vanishes for back-scattering ($\phi_i=\pi$). In a degenerate pump-probe experiment at 88 meV, i.e. well below the optical phonon threshold, the different DTS for the two orthogonal probe pulse polarizations confirm that an anisotropic distribution persists for several ps (cf. Fig. 1c). This is direct evidence that noncollinear scattering has not led to an angular thermalization. In contrast, pumping at 1.5\,eV but with almost similar pulse duration and fluence enables scattering via optical phonons (cf. Fig. 1d). As a result, the distribution probed at a photon energy of 88\,meV is isotropic for all delay times (cf. Fig. 1e). This comparison of single-color and two-color experimental results unambiguously shows that the dynamics in the vicinity of the Dirac point contains important aspects that are beyond the widely applied thermodynamic understanding of the carrier dynamics in graphene. The thermodynamic model implies thermalization on a sub-100 \,fs timescale and subsequent cooling (see also Supplemental Material). Consequently, the induced transmission on timescales longer than 100\,fs depends only on pump fluence, but not on pump photon energy. Many recent tr-ARPES and pump-probe experiments can be described very well by the thermodynamic model \cite{GierzNMat2013,Gierz2014,Mics2015,Ivanov2015}. In these experiments ultrafast thermalization is ensured by the excitation conditions, leading to both strong Coulomb and electron-phonon scattering. Our experiments show, however, that this not a general case. Comparably long timescales (few ps) for Coulomb scattering and a strong angular dependence of Coulomb scattering have also been predicted in a study on the impact-excitation of extrinsic electrons in graphene \cite{Song2013}.

Next, for low-energy excitations, the carrier dynamics is investigated in a wide range of fluences. At low fluences the induced maximum transmission change for probing the antinode states is about 2.3 times larger than for probing the node states, as shown in Figure 2a and 2e. In the absence of scattering or for purely collinear scattering, one would expect a factor of 3 \cite{Mittendorff2014}. With increasing fluence the ratio between the maximum transmission obtained for probing the antinode and node states $\Delta T_{an}/\Delta T_{n}$ asymptotically approaches the value 1, corresponding to a completely angularly thermalized distribution. Apart from the signal amplitudes, also the temporal positions of the signal maxima for the two probing conditions feature a pronounced fluence dependence (see Fig. 2a-d and 2f). The peaks for the two probe polarizations occur almost at the same time when low and high fluences are applied. At intermediate fluences ($\sim0.3\,\mu$J/cm$^{2}$) the signal probing the node states is delayed by about 1.6\,ps with respect to the signal probing the antinode states. To understand this phenomenon it is instructive to discuss three different fluence regimes: low, intermediate and high. The Coulomb scattering rate scales with the number of available scattering partners, which in turn scales with the applied pump fluence. For low fluences noncollinear Coulomb scattering is most inefficient resulting in $\Delta T_{an}/\Delta T_{n} \approx 3$. In the absence of noncollinear scattering the signal for probing the node states stems from optically excited carriers around $\Phi_{k} = \pi/4$ (and $3\pi/4$, $5\pi/4$, ... cf. Fig. 1a); consequently there is no time delay between the signals for probing the node and antinode states. At intermediate fluences noncollinear scattering becomes relevant, resulting in reduced values for $\Delta T_{an}/\Delta T_{n}$ and in a delayed maximum for probing the node states. This delay indicates the transfer of carriers from antinode states to node states. Finally, in the high-fluences regime, full thermalization occurs on timescales beyond the resolution of the experiment. This results in basically identical DTSs for probing antinode and node states, respectively. Naturally these curves feature similar amplitudes and similar temporal positions of the maxima. 

Note that saturation of pump-probe signals due to phase space filling can also affect the amplitude and temporal position of the induced transmission maxima. However, while the latter effect can qualitatively explain the reduction of the anisotropy with increasing fluence, it cannot explain the peculiar shift of the maximum for probing the node states. We have performed polarization-resolved experiments also at elevated temperatures up to room temperature (not shown). For temperatures above 100\,K the decay of the signal becomes fast and the amplitude drops rapidly (see also \cite{Winnerl2011}) resulting in a decreased signal-to-noise ratio that does not allow one to trace the noncollinear scattering contribution over a large range of fluences. Nevertheless, we could not find any indication that the dynamics of the noncollinear Coulomb scattering are influenced by the substrate temperature up to $300\,\text{K}$ (for more details, see the Supplemental Material \footnote{See Supplemental Material at [URL] for information about the temperature dependence of the noncollinear Coulomb scattering}).

For a detailed insights into the relaxation dynamics and also for a clear separation of the saturation effect from noncollinear scattering microscopic modeling is applied. The modelling is performed for an intrinsic graphene monolayer, nevertheless the results are still valid for the more complex sample structure used in the experiment. For more information see the Supplemental Material. For directly visualizing the carrier dynamics, the carrier occupations along the direction of the center of the antinodes (center of the nodes) corresponding to $\bm{k} \bot \bm{E_{pump}}$ ($\bm{k} \| \bm{E_{pump}}$) are depicted in Figure \ref{fig:theory} for two fluences in a range where the largest qualitative differences are found. At low fluences and early times the optically generated carriers at an energy of $\hbar\omega/2=\unit[44]{meV}$ are clearly visible in the occupation along $\Phi_k=\pi/2$. However, a hot distribution is reached rapidly along the respective $\Phi_{\bf k}$-direction due to efficient collinear scattering. As noncollinear scattering is less efficient, the distribution remains anisotropic on a timescale of several ps. For higher pump fluences, cf. Fig. \ref{fig:theory}d-f, both collinear and noncollinear Coulomb scattering becomes more efficient. As a result, thermalization along a particular $\Phi_{\bf k}$-direction is considerably faster and an almost isotropic distribution is reached within $\sim$4\,ps. Note that even for high fluences the occupation numbers at the probe energy of 44\,meV are well below half-filling of available states. Therefore, saturation effects can be ruled out as the main mechanism of the observed dynamics.

With increasing pump fluence the temperature of the thermalized electrons increases. Consequently, also electronic states above 200\,meV eventually become populated and thus emission of optical phonons may contribute to the relaxation dynamics \cite{Winnerl2011}. To investigate the strength of this effect, the DTS curves are calculated for the full dynamics and for the dynamics without phonons. The differences are negligible, a detailed comparison is shown in the Supplemental Material. In summary those calculations, together with the extraordinary lifetime of the anisotropy compared to experiments in the near infrared \cite{Mittendorff2014, Trushin2015} and the comparison of the two color and the degenerate pump probe experiment (cf. Fig. 2c and 2e), prove that noncollinear Coulomb scattering is the only remaining relaxation channel to reduce the anisotropy. This channel can be effectively controlled and studied by the variation of the pump fluence.

\begin{figure}[t]
\includegraphics[width=0.48\textwidth]{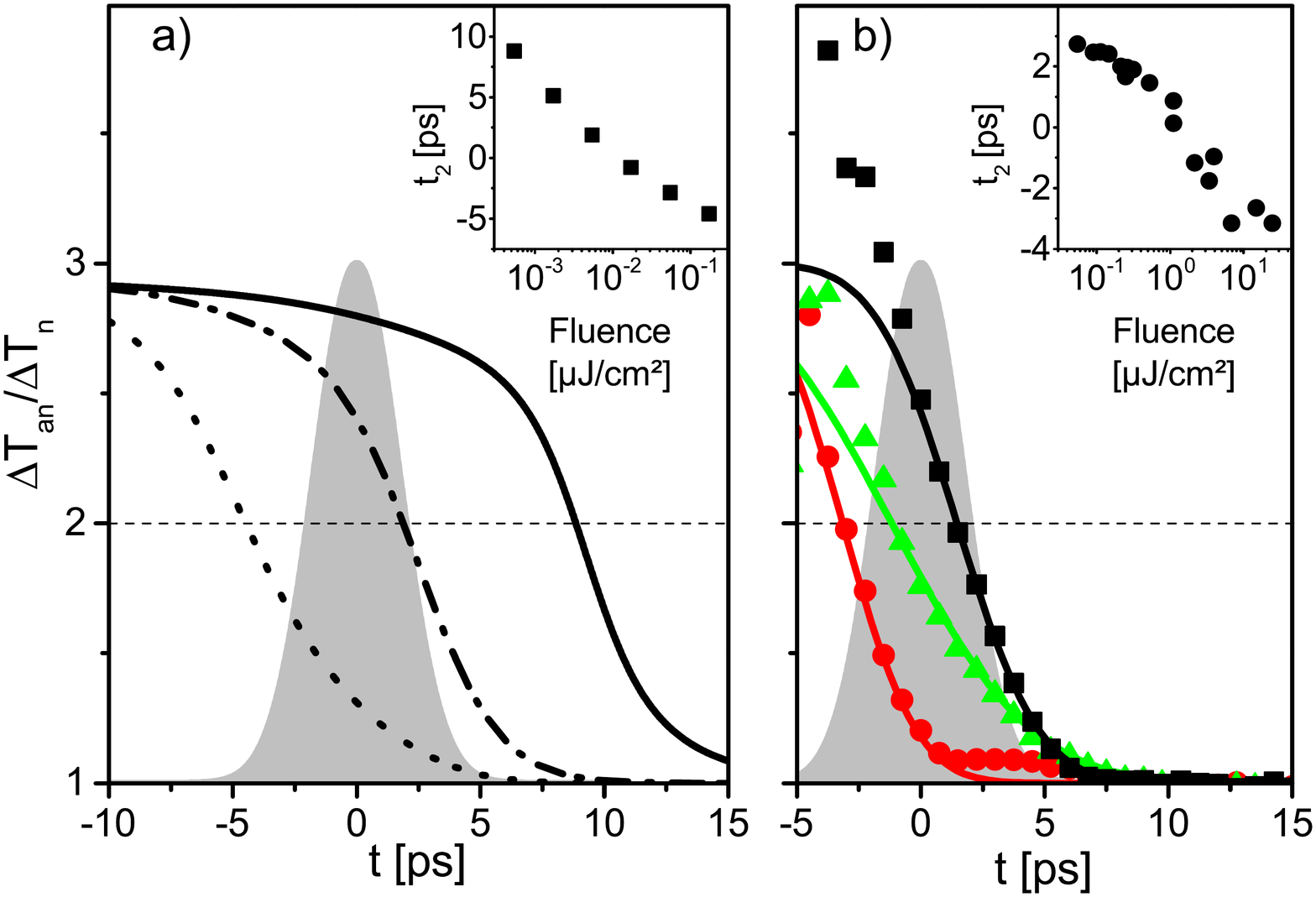}
	\caption{The decay of the ratio $\Delta T_{an}/\Delta T_{n}$ is depicted for different fluences. a) theoretical data: $\unit[171]{nJ/cm^2}$ dotted; $\unit[5.4]{nJ/cm^2}$ dot-dashed; $\unit[0.54]{nJ/cm^2}$ solid. b) experimental data: $\unit[24]{\mu J/cm^2}$ red; $\unit[2.1]{\mu J/cm^2}$ green; $\unit[0.5]{\mu J/cm^2}$ black. The gray shaded areas indicate the pump pulse. For the experimental data in b) fit curves based on an error function, that were used to extract the value of $t_{2}$, are shown. In the insets the time $t_{2}$ ascribed to the half-way decay of the optically induced anisotropy is shown over the fluence. \label{fig:anidec}}
\end{figure}

Finally, we investigate the decay of the anisotropic carrier distribution into a thermalized distribution quantitatively and compare experiment and theory. To this end, the temporal evolution of the ratio between the induced transmissions for probing the antinode and node states, respectively, $\Delta T_{an}/\Delta T_{n}$ is evaluated for both experimental and theoretical data (see Figure 4). Starting from a value of 3, corresponding to the optically excited non-equilibrium distribution, the ratio $\Delta T_{an}/\Delta T_{n}$ drops to the value of 1, characterizing the isotropic thermalized distribution. While the overall shape of this ratio as a function of time depends only weakly on fluence, the point in time at which the distribution loses its anisotropic character depends strongly on fluence. For low fluences the anisotropic distribution persists until the pump pulse has vanished, while for high fluences a significant loss in anisotropy occurs already during the rising edge of the pulse (cf. Fig. 4a). The experimental dynamics (cf. Fig. 4b) is qualitatively very similar to the theoretically predicted one. Since for early times the signal is small compared to the noise, the ratio $\Delta T_{an}/\Delta T_{n}$ at those times is instable as one essentially divides by zero. Note that a significant part of the dynamics occurs during the pump pulse, hence there is a balance of refreshing the anisotropic distribution by optical pumping and a decay of the anisotropy by noncollinear scattering. For a quantitative comparison the time $t_{2}$ at which the distribution is half-way angularly thermalized (i.e. $\Delta T_{an}/\Delta T_{n} = 2$), is extracted from both the theoretical and experimental data (cf. insets of Fig. 4a and 4b). In both experiment and theory $t_{2}$ is positive for small fluences, reflecting a pronounced anisotropic distribution after the excitation pulse maximum. Increasing noncollinear Coulomb scattering by applying higher fluences results in a reduction of $t_{2}$ down to values corresponding to the very front of the excitation pulse ($\sim$ -4\,ps). However, there is a notable offset between the theoretical and the experimental data, i.e. higher fluences are required in the experiment to achieve similar effects. This might be ascribed to an overestimation of the fluence in the experiment, since the fluence is reduced for the layers underneath the top layer. Furthermore there is an uncertainty in the determination of the temporal position of the pump-pulse maximum. Last but not least it has to be considered that no adjustable parameters are used in the comparison.

The comparably long time for angular thermalization is expected to be advantageous for infrared and THz device applications based on hot carriers such as fast detectors and modulators \cite{Sensale-Rodriguez2012,Mittendorff2015,TielrooijJPCM2015}. To improve the efficiency of such devices the higher electron temperature in the direction perpendicular to the direction of the polarization can be exploited by applying appropriate electrode geometries. The relaxation time on the ps timescale is attractive as it is long enough to result in a significant contribution to the extractable signal of such devices. On the other hand, the ps timescale is fast enough to prevent a deterioration of the operation speed of the device, resulting in a high temporal resolution. Anisotropic photoconductive effects have been observed for excitation with visible radiation \cite{Kim2012}. Our study suggests that this effect should be much larger for photon energies below the optical phonon energy. Besides its fundamental importance, the present study is of particular relevance for gated devices, where the Fermi level is shifted to the charge neutrality point \cite{Tong2015,Koppens2014}. Furthermore our calculations predict that an anisotropic carrier distribution on a ps timescale can be also observed in doped graphene, however in this case the pump-induced change in transmission is much smaller (see Supplemental Material). These preliminary results suggest that our main findings may be also relevant for devices employing doped graphene. For all the mentioned device applications it is highly attractive that anisotropic carrier distribution persists also at room temperature.
\FloatBarrier

In conclusion, our study has revealed that low energy Coulomb scattering in almost intrinsic graphene exhibits an unusual twofold nature: It rapidly thermalizes the distribution along all $\bm{k}$-directions pointing radially
away from the Dirac point, while preserving the optically induced anisotropy on a ps timescale. This effect is attractive for optoelectronic devices based on hot carriers.

\appendix*
\begin{acknowledgments}
We thank P. Michel and the ELBE-team for their dedicated support. This work was financial supported by the SPP 1459 (S.W., E.M.) and SfB 910 (A.K.) of the DFG. E.M. and C.B. also acknowledge financial support from the EU Graphene Flagship and E.M. from the Swedish Research Council (no. 604391). W.d.H. acknowledges support from the AFSOR. We thank H. Mittenzwey, who acknowledges support from GRK 1558, for his help.
\end{acknowledgments}
\bibliography{Noncollinear-Coulomb-scattering}
\end{document}